\documentclass[a4paper,fleqn,usenatbib]{mnras}

\usepackage{newtxtext,newtxmath}

\usepackage[T1]{fontenc}
\usepackage{ae,aecompl}

\usepackage{graphicx}	
\usepackage{amsmath}	
\usepackage{amssymb}	
\usepackage{hyperref}       

\def\CN2{\mbox{$C_N^2 $}}
\def\CT2{\mbox{$C_T^2$}}
\def\tauO{\mbox{$\tau_{0} $}}

\title[Towards an automatic wind profiler for WFAO systems]{Towards an automatic wind speed and direction profiler for Wide Field AO systems}
\author[G.Sivo et al.]{
G. Sivo,$^{1}$\thanks{E-mail: gsivo@gemini.edu}
A. Turchi,$^{2}$
E. Masciadri,$^{2}$\thanks{E-mail: masciadri@arcetri.astro.it}
A. Guesalaga,$^{3}$
and B. Neichel$^{4}$
\\
$^{1}$  Gemini Observatory, AURA, Colina el Pino s/n, Casila 603, La Serena, Chile\\
$^{2}$ INAF Osservatorio Astrofisico di Arcetri, Largo Enrico Fermi 5, I-501 25 Florence, Italy\\
$^{3}$ Pontificia Universidad Catolica de Chile, 4860 Vicuna Mackenna, Casilla 7820436, Santiago, Chile\\
$^{4}$ Aix Marseille Universit\'e, CNRS, LAM Laboratoire d'Astrophysique de Marseille UMR 7326, 13388, Marseille, France}

\date{Accepted 2018  20. Received 2017 December 20; in original form 2017 September 26}

\pubyear{2018}

\begin{document}
\label{firstpage}
\pagerange{\pageref{firstpage}--\pageref{lastpage}}
\maketitle

\begin{abstract}
Wide Field Adaptive Optics (WFAO) systems are among the most sophisticated AO systems available today on large telescopes. The knowledge of the vertical spatio-temporal distribution of the wind speed (WS) and direction (WD) are fundamental to optimize the performance of such systems. 
Previous studies already proved that the Gemini Multi-Conjugated AO system (GeMS) is able to retrieve measurements of the WS and WD stratification using the SLODAR technique and to store measurements in the telemetry data. In order to assess the reliability of these estimates and of the SLODAR technique applied to such a kind of complex  AO systems, in this study we compared WS and WD retrieved from GeMS with those obtained with the atmospherical model Meso-Nh on a rich statistical sample of nights. It has been previously proved that, the latter technique, provided an excellent agreement with a large sample of radiosoundings both, in statistical terms and on individual flights. It can be considered, therefore, as an independent reference. The excellent agreement between GeMS measurements and the model that we find in this study, proves the robustness of the SLODAR approach. To by-pass the complex procedures necessary to achieve automatic measurements of the wind with GeMS, we propose a simple automatic method to monitor nightly WS and WD using the Meso-Nh model estimates. Such a method can be applied to whatever present or new generation facilities supported by WFAO systems. The interest of this study is, therefore, well beyond the optimization of GeMS performance.
\end{abstract}



%
%
\section{Introduction}
Wide Field Adaptive Optics (WFAO) systems are among the most technologically advanced and complex AO systems installed or to be installed on large telescopes. This term includes a few types of adaptive optics: the laser tomography adaptive optics (LTAO), the multi-conjugated adaptive optics (MCAO), the ground layer adaptive optics (GLAO) and the multi-object adaptive optics (MOAO). All the WFAO systems have the common characteristics to increase significantly the field of view (FoV) of the images obtained after the AO correction and the portion of sky that become accessible after the reconstruction of the original wavefront. Indeed, WFAO can obtain a corrected wavefront on a field of view of the order of a few arc-minutes. This overcomes the order of a few arc-seconds typical of the single conjugated adaptive optics system (SCAO). In WFAO systems, light from a set of guide stars (GSs) located in dedicated configurations on sky, is used to probe the instantaneous 3D phase perturbations, and an inverse problem procedure is used to reconstruct the turbulence in the sensed volume. This technique, is known as atmospheric tomography. \cite{tallon1990} proposed for the first time this technique and, later on, some improvement have been suggested by several other authors \citep{johnston1994,ellerbroek1994,fusco2001}. 

The knowledge of the three-dimensional turbulence distribution as well as of the wind speed (WS) and wind direction (WD) stratification on the whole atmosphere strongly affect performances of all WFAO systems. Tomographic performance can be deeply deteriorated by wrong quantification of the atmospheric state \citep{neichel2008}. The geometry of the system is strictly correlated to some important limitations (among other the modes and the turbulence that is not seen by the system). The regularisation of these modes is affected by errors on the identification of height of the recovered turbulent layers. The sensitivity of the reconstructors to unseen frequencies is strictly related to the height of the recovered layers that are seen by the AO system. A wrong geometry implies that the regularisation acts in an inefficient way. Different papers report analyses on how errors of the model affect the turbulence quantification: \cite{conan2001} and \cite{tokovinin2001} investigated how the reconstructed error is influenced by an error on the estimate of the \CN2 profile; \cite{lelouarn2002} studied which was the relationship between the reconstruction of a limited number of turbulent layers at discrete altitudes and the whole volume of atmospheric turbulence; \cite{fusco1999} explored the impact on the reconstruction of the wavefront by employing a small number of equivalent layers. Conclusions indicated that a small number of layers (2-3) was sufficient to reconstruct a uniform corrected field; more recently, however, \cite{costille2012} concluded that, for an application to ELTs, the sampling of the \CN2 profiles (that is the ability to reconstruct thin turbulence layers) plays an important role on the performance of the system and the number of turbulent layers necessary to the reconstruction of the wavefront can be higher than this. All those previous investigations tell us that the FoV is directly proportional to the sensitivity of the AO system with respect to errors that can be done in identifying the height and position of the layers. Beside to the \CN2 profiles a good WS and WD estimate is determinant for an efficient employment of an MCAO system. Firstly, because the WS, together with the \CN2, determines the value of the wavefront coherence time ($\tau_{0}$) that tells us how fast an AO system has to run. Secondly, because errors in time in an adaptive optics are produced by the modification of the turbulence between the instant in which a perturbed wavefront is sensed and the instant in which is corrected. This is known as 'control cycle'. 
If the turbulence is assumed to be frozen, the forecast of the state of turbulent layer at a time in the future depends on the state of the WS and WD of the atmospheric turbulent layer at present \citep{johnson2008}. By using the CANARY demonstrator at the William Hershel Telescope, it has been demonstrated, on-sky, that a predictive control, based on a Linear Quadratic Gaussian (LQG) controller fed by atmospheric turbulence priors such as \CN2 and wind speed profiles (only wind speed during that experience), improved significantly the system performance with respect to a classic integral controller. It is shown that, in classical AO, the improvement in Strehl Ratio is of the order of 10\% in K-band in average \citep{sivo2014}. This predictive control has also been validated in WFAO (MOAO configuration). It has been shown as well that, by using an LQG control, better performance are obtained than by using a standard Minimum Mean Square Error (MMSE) reconstructor. The gain in performance is clear when the wind speed profile, given as {\it 'a priori'} to the model, is close to the real one \citep{sivo2013,osborn2015}. Recently, also \cite{ono2016} dealt about the advantages one can obtain in knowing the wind speed in applications to  the wavefront reconstruction in MOAO systems conceived for ELTs.
The fact to know the WS and WD of the different turbulent layers is, therefore, very important to reconstruct the state of the phase in the time scale of the control cycle as other authors stated \citep{gavel2002,poyneer2007,ammons2012}. The last paper also put in evidence the potential additional benefit to tomographic AO systems in knowing the WS: the combination of wind speed and phase height information from multiple guide stars breaks inherent degeneracies in volumetric tomographic reconstruction, producing a reduction in the geometric tomographic error.

\begin{figure*}
\begin{center}
\includegraphics[width = 0.95\textwidth]{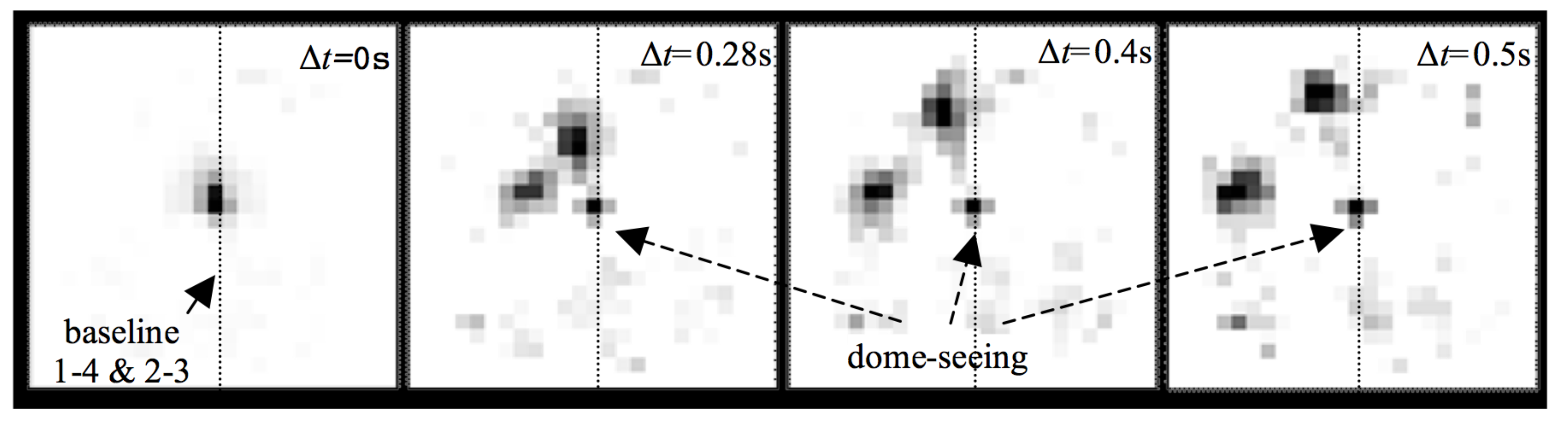}
\end{center}
\caption{\label{fig:movie} The sequence reported in the figure shows the correlation time from $\Delta$t = 0 s to $\Delta$t = 0.5 s, with two layers moving in different directions. In the centre of each individual square is visible a third static peak, corresponding to the dome seeing.}
\end{figure*} 

In this paper we perform a comparison between wind speed reconstructed by the mesoscale non-hydrostatical atmospherical model Meso-Nh and measured by GeMS on a sample of 43 nights, very representative from a statistical point of view. Previous studies \citep{masciadri2013,masciadri2015} proved that the Meso-Nh model could reconstruct reliable WS and WD profiles on the total atmosphere (from 0 to 20~km above the ground) above one of the best astronomical sites in the world (Cerro Paranal in Chile). Comparison of model predictions with 50 radio-soundings launched in winter and summer time of the same year have been studied in statistical terms. For the wind speed, it has been found a bias within 1~ms$^{-1}$ and a RMSE within 3 ms$^{-1}$ in the range 5-15~km. In the 3-5~km range and above 15~km, the bias is within 2 ms$^{-1}$. Also a dedicated study performed on all the 50 couples (radiosoundings and model profiles) (Masciadri et al. 2013 - Fig.B1) told us that a similar satisfactory performance were obtained by the model, in statistical terms and also by comparing radiosoundings vs. model ouputs on each individual couple. It has been possible to conclude, therefore, that the Meso-Nh model could be taken as a reference to validate the WS and WD provided by GeMS. The distance between Cerro Paranal and Cerro Pach\'on is indeed of the order of 600-700 kilometers and they are located in the same mountain Chilean region. There is, therefore, no reason to assume that the model can provide different reliability in its estimates in the two sites for the vertical stratification of the WS on the whole 20~km.

The goal of this paper is to validate the reliability of GeMS estimates as a profiler of WS and WD and, equally, to investigate if it possible to use the Meso-Nh to feed GeMS to obtain an efficient automatic WS and WD profiler applied to WFAO systems. Indeed, as it will be described later on, the procedure to automate GeMS measurements is particularly complex and it appears more suitable to use the WS and WD estimates from the Meso-Nh model that offers the advantage of being a technique more complete in time (a temporal frequency of two minutes during the whole night vs. a limited number of detections as obtained by GeMS as it will be described later on). The Meso-Nh model offers also a better spatial coverage i.e. 62 vertical levels instead of the typical 2-3 layers detected by GeMS. Besides, it is important to remind that, the important information for a WFAO system, is the detection of the wind speed where there are turbulent layers. 
It is worth to say that it has been observed that, rapid changes of atmospheric conditions including WS and WD, determines negative effects on the quality of observations taken with GeMS. 
In the operational phase is extremely useful to know the state of the WS and WD as measured by GeMS. It should be even preferable to know in advance the wind speed with the Meso-Nh model. The fact to know the WD and WS with some anticipation might be used, for example, to update the control matrices, with an increment or a reduction of the relative gains related to each turbulent layers height if it is joint to the knowledge of the $\CN2$. The fact to know the WS and WD is extremely useful in the context of predictive control, for example for tip-tilt control \citep{sivo2014,juvenal2016}. 
%
%
\section{Observations}
\label{sec_obs}

GeMS, the MCAO facility installed at Gemini South, is currently the first LGS-based MCAO system using sodium stars in regular operation dedicated for astronomic observations \citep{rigaut2014,neichel2014}. It employs five laser guide stars (LGSs) projected on a 1 arcmin squared asterism (four at the corners and one in the middle). GeMS uses this artificial asterism to measure and correct for atmospheric perturbations and provides an almost diffraction limit corrected image in the near-infrared (NIR) over a wide FoV of about 2 arcminutes. GeMS is currently feeding three scientific instruments: a 4k $\times$ 4k NIR imager called GSAOI \citep{mcgregor2004}, a wide FoV NIR imager and spectrograph Flamingos2 \citep{elston2003}, and more recently a visible multi-object spectrograph and imager GMOS-S \citep{crampton2000,hibon2014,hibon2016}.

We retrieve turbulence and WS profiles employing a procedure presented in \cite{cortes2012} that uses the SLODAR technique applied to binaries \citep{wilson2002} and adapted to multiple LGSs available on WFAO instruments working with Shack-Hartmann wave-front sensors \citep{cortes2012,gilles2010,osborn2012}. 
Turbulence strength has been sampled with a vertical resolution equal to $\Delta$h = d/$\theta$ where d is the sub-aperture of the wavefront sensor projected on the pupil and $\theta$ the angular distance of different couples of stars chosen among the LGSs constellation of GeMS. Depending on the selected binaries, we can obtain, therefore, different $\Delta$h. 
Since we use LGSs, we need to deal with a well-known limitation called the {\it cone effect}. This implies that the bins along z-axis have a different size (see Eq.1 in \cite{cortes2012}).  

As an extended version of the SLODAR technique, we determine the wind vertical {\it profiling} by computing cross-correlations with different delay times between all the different vectors of arrival angles i.e. slope measurements from the valid WFS sub-apertures \citep{wang2008}. 
This profiling technique provides information on how the turbulence is distributed in the atmosphere and on the temporal evolution of the WS and WD of all identified layers. In case of GeMS, since the LGSs are centre launched, the slopes suffer from the fratricide effect. The method has been modified as well to tackle this issue \citep{guesalaga2014}. 

The selection of observations has been done manually in this analysis. That means that GeMS measurements have been selected in a way that cross-correlation peaks (related to turbulent layers moving rigidly in the atmosphere) are clearly detected without ambiguity. To detect the WS of a layer it is required to follow the displacement of the peak in the cross-correlation map. These peaks are more easily recognised when the atmosphere is characterised by turbulent layers placed at different heights with different velocity that produce peaks with relative tracks that do not overlap among them. Data analysis is performed following the procedures developed in \cite{cortes2012} and \cite{guesalaga2014}. For each telemetry file, a cross-correlation temporal movie is built. Fig.\ref{fig:movie} shows a sequence of images selected from one of this temporal movies. The WS is retrieced from the displacement of the correlation peak over a given $\Delta$t.
For the WS uncertainties we assumed a $\pm$0.2 to $\pm$ 0.5 sub-aperture error depending on the signal to noise ratio of the peak, and we assumed $\pm$ 1 to $\pm$ 2 frames error depending on the shift velocity of the peak. Uncertainties on the estimation of heights is based on the size of altitude bin, as reported in \cite{cortes2012}. As it is reported in Section \ref{wind_speed_sec} the uncertainty along the z-axis is of the order of $\pm$ [0.5, 0.8] km. 

\clearpage
%
%
\section{Model} 

We performed the numerical simulations on the selected 43 nights sample using Meso-Nh \citep{lafore1998}, which is a non-hydrostatic mesoscale atmospheric model developed by the Centre National des Recherches M\'et\'eorologiques (CNRM) and the Laboratoire d'A\'ereologie (LA) of Universit\'e Paul Sabatier (Toulouse). As all mesoscale model, Meso-Nh provides spatio-temporal evolution of meteorologic parameters, on a 3D spatial scale, over a finite region of the Earth. The anelastic formulation of the set of hydrodynamic equations allows for the filtering of acoustic waves. The physical exchange between soil and atmosphere is computed by making use of ISBA (Interaction Soil Biosphere Atmosphere) module \citep{noilhan89}. The model uses 
the \cite{gal75} coordinates system on the z-axis and the C-grid (in the formulation of \cite{arakawa76}) for the spatial digitalization. The time evolution scheme use a custom three-steps time-filtered leap-frog as described in \citep{asselin72}. The turbulence model is a one-dimensional 1.5 turbulence closure scheme \citep{cuxart00}, with a one-dimensional mixing length first described by \cite{bougeault89}. From the basic turbulence scheme, the optical turbulence parameters ($C_{N}^{2}$ and derived integrated parameters) are computed with the Astro-Meso-Nh package developed by \cite{masciadri1999a}. The last package has seen a great amount of development in recent years supporting many studies that addressed the reliability of this forecast method in astronomical applications. The most recent version of the Astro-Meso-Nh code has been described in \cite{masciadri2017}. 
The Meso-Nh and the Astro-Meso-Nh models have been parallelized with OPEN-MPI with a great degree of scalability, allowing users to make use of multi-core workstations, local clusters or even large High Performance Computing Facilities (HPCF), in our case the HPC of the European Centre for Medium Range Weather Forecasts (ECMWF) and achieving, under specific model configurations, a smaller computing time. 

\begin{figure}
\begin{center}
\includegraphics[width = 1.0\linewidth]{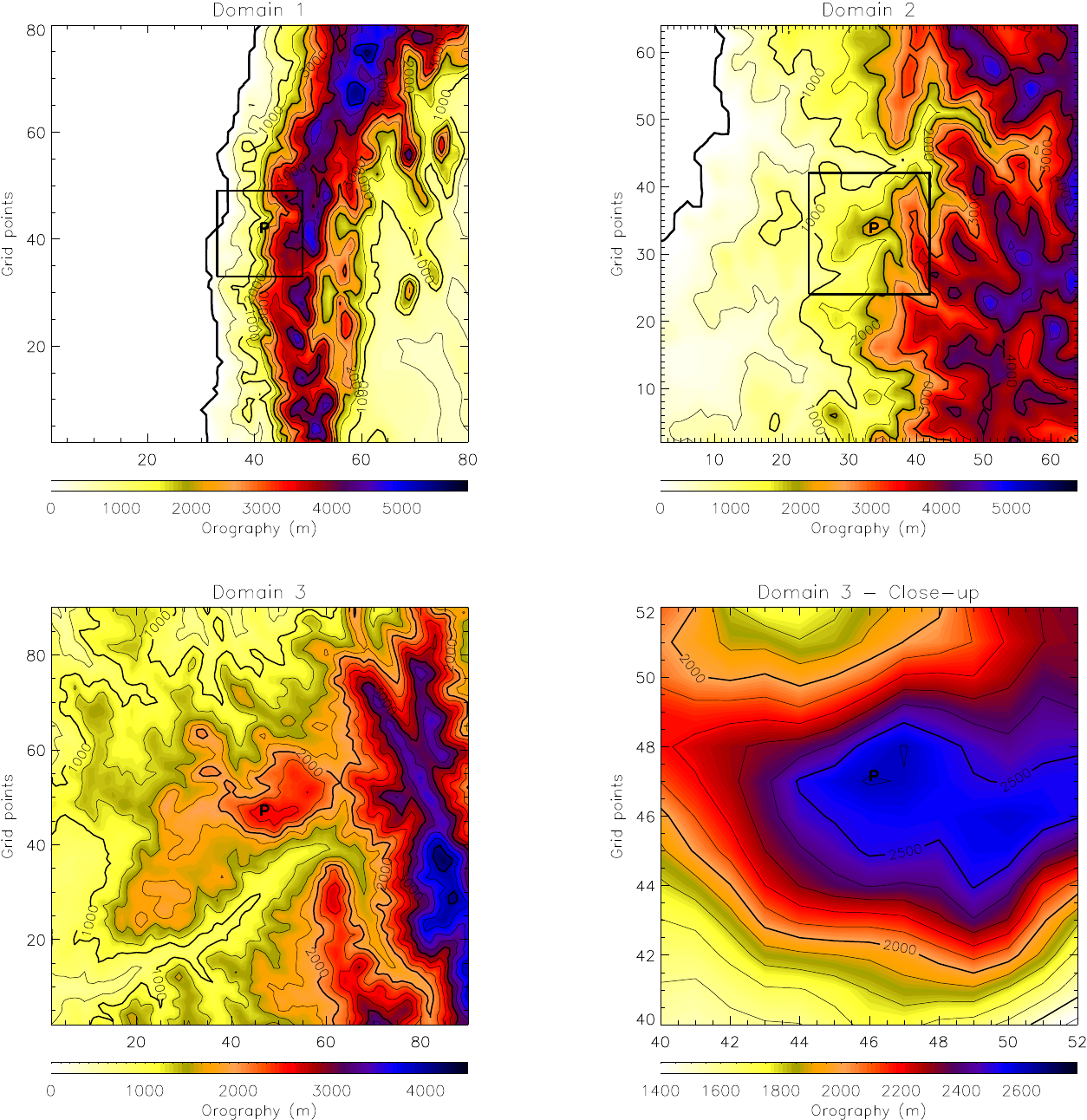}
\end{center}
\caption{ \label{fig:orography} Topography. Domain 1 (top-left): 80x80 grid points, 100kmx100km for $\Delta$(X)=10km; Domain 2 (top-right): 64x64 grid points, 160km$\times$160km for $\Delta$(X)=2.5km; Domain 3 (bottom-left): 90$\times$90 grid points, 45km$\times$45km for $\Delta$(X)=0.5 km. A zoom of the summit extracted from Domain 3 is shown on the bottom-right.}
\end{figure} 

\begin{table}
\caption{Grid-nesting configuration of the Meso-Nh model.
The number of horizontal grid-points is reported in the second column, the domain extension for each imbricated domain in the third column and
the horizontal resolution $\Delta$X in the fourth column.}
\label{tab:orog}
\begin{center}
\begin{tabular}{|c|c|c|c|}
\hline
Domain & Grid   & Domain size & $\Delta$X \\
       & Points & (km)        &  (km)  \\
\hline
Domain 1 &  80$\times$80  & 800$\times$800 & $\Delta$X = 10   \\
Domain 2 &  64$\times$64  & 160$\times$160 & $\Delta$X = 2.5 \\
Domain 3 & 90$\times$90 &  45$\times$45  & $\Delta$X = 0.5   \\
\hline
\end{tabular}
\end{center}
\end{table}

Simulations have been performed above the site of GeMS i.e. Cerro Pach\'on at (70$^{\circ}$ 44' 12.096'' W; 30$^{\circ}$ 14' 26.700" S)\footnote{\href{http://www.gemini.edu/sciops/telescopes-and-sites/locations}{http://www.gemini.edu/sciops/telescopes-and-sites/locations}}. 

To maximize the resolution of the simulations over the site of interest, we used the so-called grid-nesting' technique, which consists of producing multiple  (three in this case) imbricated model domains, each progressively extended on smaller surfaces and increased horizontal resolution and centered on the Cerro Pach\'on coordinates, up to a maximum resolution of the innermost domain $\Delta$X~=~500~m (see Fig.\ref{fig:orography} and Table\ref{tab:orog}). The vertical grid remains identical in each domain. It has already been proven in a previous studies \citep{masciadri2013} that this model configuration, together with the detailed 500~m resolution of the innermost domain, allows to obtain excellent results in terms of the vertical stratification of wind speed and direction, which are the parameters of interest in this study.
The Digital Elevation Model (DEM, i.e. topography) used for the most external domains (1 and 2) is the GTOPO30\footnote{\href{https://lta.cr.usgs.gov/GTOPO30}{https://lta.cr.usgs.gov/GTOPO30}}, which have an intrinsic horizontal resolution of 1~km. In domain 3 we used the Shuttle Radar Topography Mission (SRTM)\footnote{\href{http://srtm.csi.cgiar.org/}{http://srtm.csi.cgiar.org/}} topography, which has an intrinsic horizontal resolution of 90~m. The model interpolates the Innermost DEM using the specific horizontal resolution of 500~m as reported in Table \ref{tab:orog}. In Fig.\ref{fig:orography} we show a graphical representation of the three imbricated domains.

We used a vertical grid of 62 levels which covers the whole atmosphere up to $\sim$ 20~km above the ground level (a.g.l.), starting from an initial point at 5~m a.g.l. with a progressive logarithmic stretching of 20$\%$ up to 3.5~km a.g.l.. From this last height onward the model uses an almost constant grid size of 600~m. 
Even if the number of model levels can be freely chosen and it is not a constraint for the model, it is our interest to limit the number of vertical model levels to what it is necessary to avoid to increase the computation time. We note that there are no turbulent layers (i.e. no WS measurements) detected by GeMS at heights higher than 18~km a.s.l. The configuration chosen is therefore suitable for this study. Besides that, we highlight that this configuration is also suitable to estimate the $\tauO$ \citep{masciadri2017}. Indeed, if we look at the climatology of the $\CN2$, at latitudes typical of astronomical observatories, we observe that the optical turbulence is very weak above 20~km (for example \cite{garcia2011},\cite{lor2011b}). If we look at the climatology of the WS, we observe that the wind speed reaches a maximum at 11-13 km and then it decreases inexorably at heights of the order of 20~km or more (see for example \cite{hagelin10}). This is therefore a reasonable spatial scale.


We performed the numerical simulations on a total of 43 nights for which we have GeMS telemetry data available. For each night, identified by the UT date, the simulation was initialized at 18:00 UT of the previous day and new updated {\it forcing} data was fed to the simulation every 6 hours, using analyses input data produced in the meanwhile by the Global Circulation Model of the European Centre for Medium-Range Weather Forecasts (ECMWF). Each simulation lasted up to 11:00 UT (07:00 LT), for a total simulation length of 17 hours. The temporal sampling of the model outputs i.e. vertical wind profiles is 2 minutes, however the analysis performed in this paper made use only of the specific data referring to local night times for those selected times in which observations/measurements are available.

%
%
\section{Results} 
\subsection{Wind speed}
\label{wind_speed_sec}

\begin{figure*} 
\begin{center}
\includegraphics[width=0.85\linewidth]{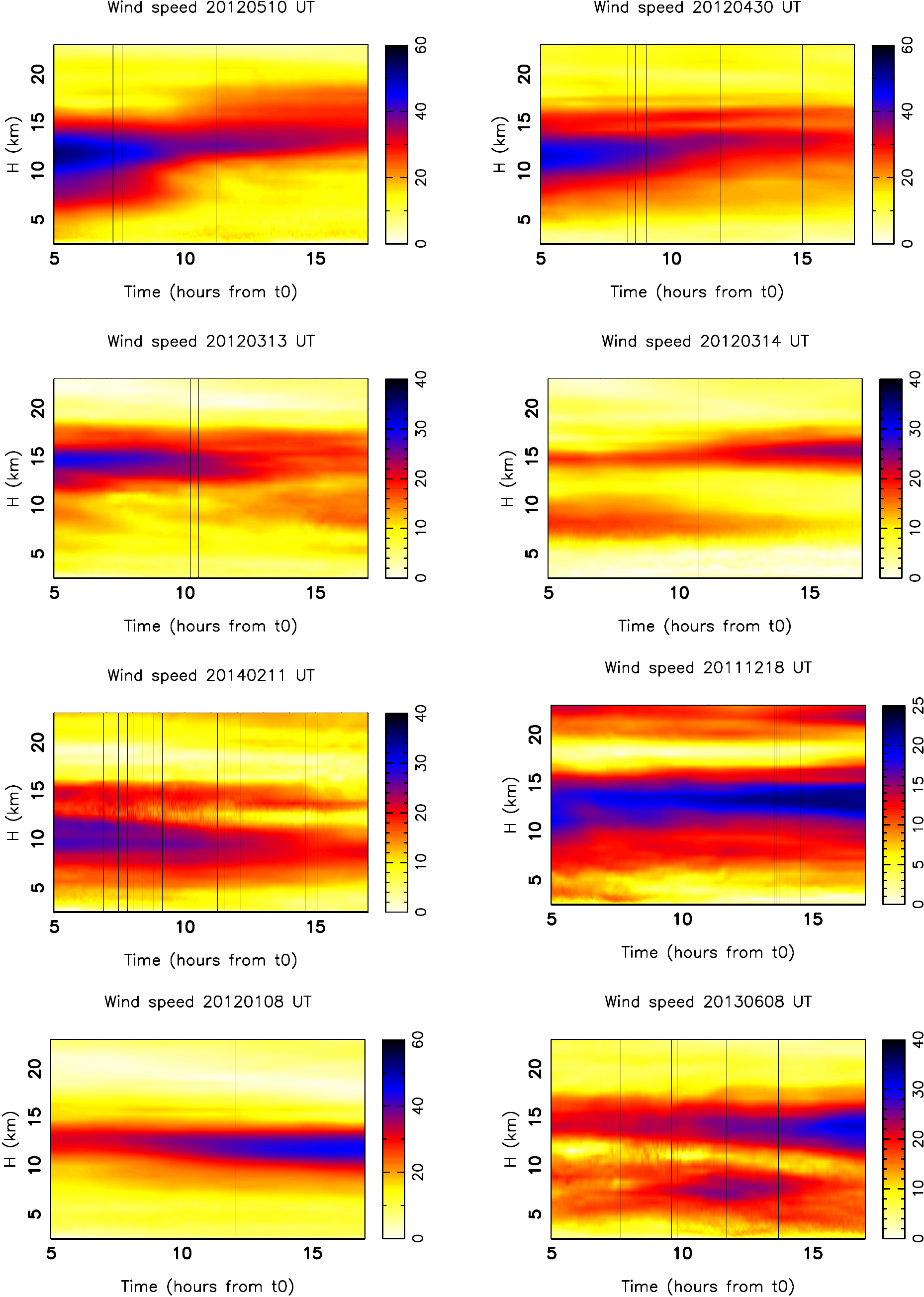}
\end{center}
\caption{\label{fig:temp_evol1} Temporal evolution of the WS on the whole atmosphere 20~km as recovered by the Meso-Nh model during a few nights. The wind speed expressed in ms$^{-1}$ is reported in the color legend. Time is displayed on the x-axis starting from 5h after the beginning of the simulation (19:00 LT). Simulations end 17h after the beginning of the simulation (07:00 LT). The figure displays, therefore, the night time. Black vertical lines indicate the time in which measurements from GeMS are available. }
\end{figure*}

To guarantee a fair analysis the data reduction of GeMS measurements and the Meso-Nh simulations related to the whole sample of 43 nights have been done in a totally independent way. To have an idea of the features of the WS reconstructed by the model, Fig.\ref{fig:temp_evol1} reports the temporal evolution of the WS vertical profiles during the whole night for a sub-sample of nights. The black vertical lines correspond to the time in which we have WS measurements from GeMS. At each instant GeMS detects WS measurements associated to a finite number of layers (typically one to three layers). We mean with that the maximum number of layers retrieved from the technique we have described is not more than around three layers. This does not mean that in the atmosphere there are just three layers but that the method is able to discern three layers. This is due to the vertical resolution of the system but also to a set of reasons that will be described in Section \ref{discus}. Looking at Fig.\ref{fig:temp_evol1} it appears evident that, for each night, the number of estimates from GeMS is smaller than the available estimates from the model but in this context we are interested on the evaluation of the method performed by GeMS therefore this element is not critical. The important thing is to have a rich statistic of couples [observations, model outputs] to be treated.

\begin{figure*}
\begin{center}
\includegraphics[width = 0.85\linewidth]{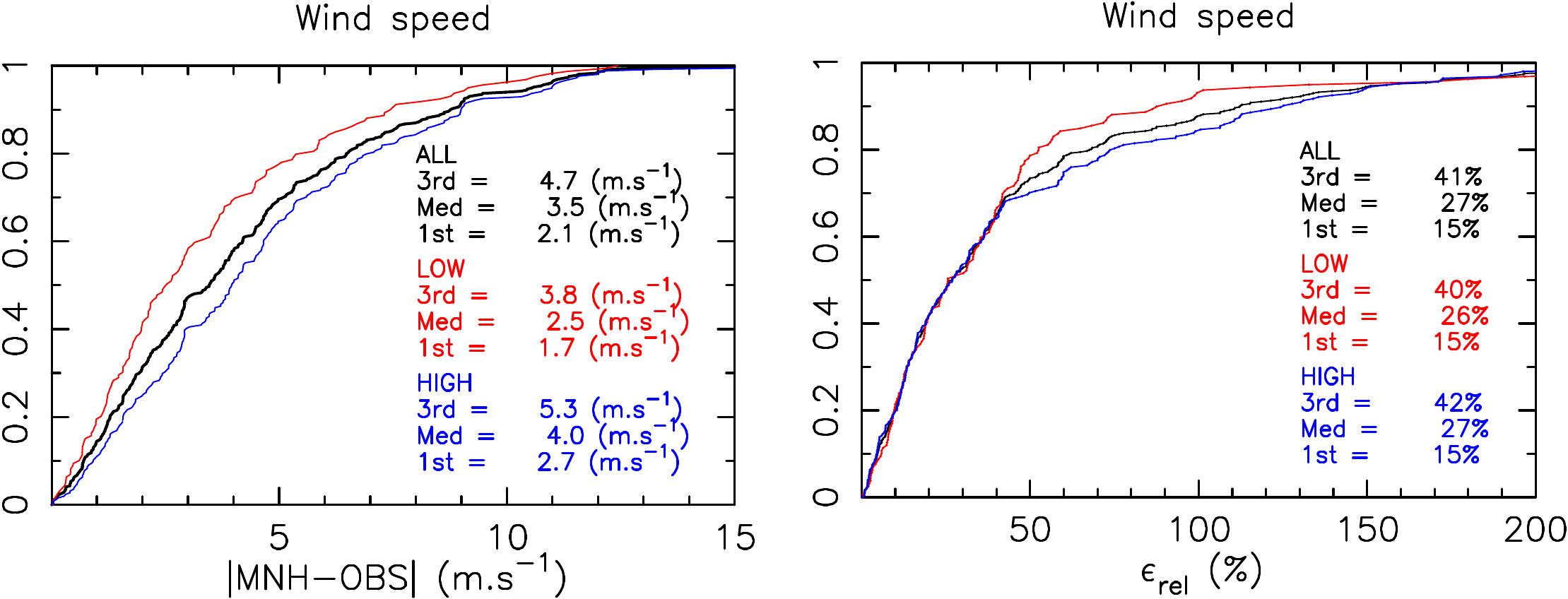}
\end{center}
\caption{\label{fig:cum_dist} Left: Cumulative distribution of the absolute difference of WS estimated by GeMS and by the model (400 points in the sample). Sample contains estimations referring to the [5~km, 18~km] a.s.l. range i.e. the high part of the atmosphere (blue line), the [3~km, 5~km] a.s.l. range i.e. the low part of the atmosphere (red line) and the [3~km, 18~km] a.s.l. range i.e. the total atmosphere (black line). Right: Cumulative distribution of the relative error of the WS in the same vertical slabs. There are no GeMS estimations for heights h > 18~km a.s.l.}
\end{figure*}

We compared model outputs with measurements on the statistical sample of 43 nights that corresponds to raround 400 couples of [observations, model outputs] estimates. Fig.\ref{fig:cum_dist} shows the cumulative distribution of the absolute difference of the WS quantified by the model and by GeMS (as Eq.\ref{eq1}) with the associated error (as Eq.\ref{eq2}) in three different vertical regions of the atmosphere: in the high part of the atmosphere [5~km, 18~km], in the low part of the atmosphere [3~km, 5~km] and in the total atmosphere [3~km, 18~km].

\begin{equation}
\left | V_{MNH}- V_{OBS}\right |
\label{eq1}
\end{equation}

\begin{equation}
\varepsilon_{rel} = \frac{\left | V_{MNH}-V_{OBS} \right |}{V_{MNH}} \times100\%
\label{eq2}
\end{equation}

There are no GeMS measurements of WS at heights higher than 18~km. All heights are above the sea level (a.s.l.). We performed the comparison using this procedure: (1) we take the measurements of GeMS $\pm$ the error bar along z-axis; (2) the differences between model and GeMS estimates is calculated inside this vertical slab. Fig.\ref{fig:cum_dist} (right side) reports the median values of 26\%, 27\% and 27\% for the relative errors respectively in the low, high and total part of the atmosphere. We obtained similar results by calculating the relative error with respect to the WS of GeMS instead of the WS of the model (this consists on replacing the respective values at the denominator in Eq.\ref{eq2}). The median value of the difference | V$_{OBS}$ -V$_{MNH}$ | in the low, high and total atmosphere is respectively 2.5 m$s^{-1}$, 4 m$s^{-1}$ and 3.5 m$s^{-1}$ as shown in Fig.\ref{fig:cum_dist} (left side).
We note that, in the first vertical grid point, model values are taken starting from 3~km a.s.l. (therefore 400~m above the ground) so to use the model in more or less the same part of the atmosphere in which it has been validated by comparison with radiosoundings\footnote{In \cite{masciadri2013} we considered 500~m instead of 400~m because of a slightly difference in the altitude of Cerro Paranal and Cerro Pach\'on. The definition of the grey zone is qualitative therefore 500~m or 400~m does not produce a great difference.}. Between 30~m and 400-500~m (what we called {\it grey zone}) it is meaningless to compare Meso-Nh and radio-soundings estimates because of topographic effects (see \cite{masciadri2013} for an extended discussion) and we can retrieve no conclusions. On the other side the model behavior close to the surface (in the first 30~m above the ground) has been validated by comparing simulations with measurements provided by a meteorological station \citep{lascaux2013,lascaux2015}. Those studies provided a very good correlations between model and measurements done by sensors located at different heights. This guarantees that the model can be considered a good reference for the surface layer too. In this analysis we observed that GeMS observations never fall in the grey zone. We conclude, therefore, that the model vs. GeMS comparison is done in a consistent way in the region where the model has been previously validated \citep{masciadri2013}.

\begin{table*}
\caption{Bias, RMSE and $\sigma$ between the model and GeMS estimations calculated in low and high part of the atmosphere on the whole sample of 43 nights. Bias is |MNH-OBS|. $^{(*)}$Referring to \citep{masciadri2013}}
\label{tab:res1}
\begin{center}
\begin{tabular}{l|ccc|ccc}
\hline
\multicolumn{1}{c}{} & \multicolumn{3}{c}{GeMS vs. Meso-Nh}&\multicolumn{3}{c}{Meso-Nh vs. Radio-soundings$^{(*)}$} \\
\multicolumn{1}{c}{} & \multicolumn{3}{c}{43 nights}&\multicolumn{3}{c}{50 nights} \\
\multicolumn{1}{c}{} & \multicolumn{3}{c}{ms$^{-1}$}&\multicolumn{3}{c}{ms$^{-1}$} \\
\hline
 Vertical slab  & bias & RMSE & $\sigma$ & bias & RMSE &$\sigma$ \\
\hline
   LOW: [3,5]~km a.s.l.& 0.47 & 4.36 & 4.33 &  -1 & 3   & 2.83 \\
   HIGH: [5,18]~km a.s.l. & -1.49 & 5.15 & 4.9 & 0.6 & 3.5 & 3.45 \\
\hline
\end{tabular}
\end{center}
\end{table*}

The bias, RMSE and $\sigma$ of the data-set in the whole atmosphere, in the high part [5~km, 18~km] a.s.l. and in the low part [3~km, 5~km] a.s.l. is shown in Table \ref{tab:res1}. $\sigma$ is the bias-corrected RMSE (Eq.\ref{eq:sigma}). 

\begin{equation}\begin{split}
\sigma=\sqrt{RMSE^2-BIAS^2}
\label{eq:sigma}
\end{split}\end{equation}

In order to provide a more complete discussion of results, the corresponding values related to the comparison between the Meso-NH estimations and 50 radio-soundings \citep{masciadri2013} are reported in the same table. These last values refer to the study in which it was proven that Meso-Nh is a reliable method for the estimation the WS and WD i.e. the study that guarantees us to consider Meso-Nh as a reference in this study. It is possible to note that bias, RMSE and $\sigma$ between the Meso-Nh model and GeMS are in a very good agreement. The statistical operators are just only slightly larger than those estimated in the \cite{masciadri2013} study. The difference for $\sigma$ is of the order of 1.5 ms$^{-1}$ in the low as well as high part of the atmosphere. However in this study we had to compare GeMS and model outputs within a larger $\Delta{h}$ than that considered in \cite{masciadri2013}. Basically a $\Delta{h}$ correspondent to the different resolution of GeMS. It is, therefore, not surprising to obtain a slightly larger error. It does not mean that results are worse. It simply indicates that we have to consider a larger uncertainty in the measurements due to the resolution but the principle of detection works properly. Considering the strong shear of the wind at this heights we consider therefore that the agreement between Meso-Nh and GeMS is very satisfactory. 

\begin{figure*} 
\begin{center}
\includegraphics[width=0.9\linewidth]{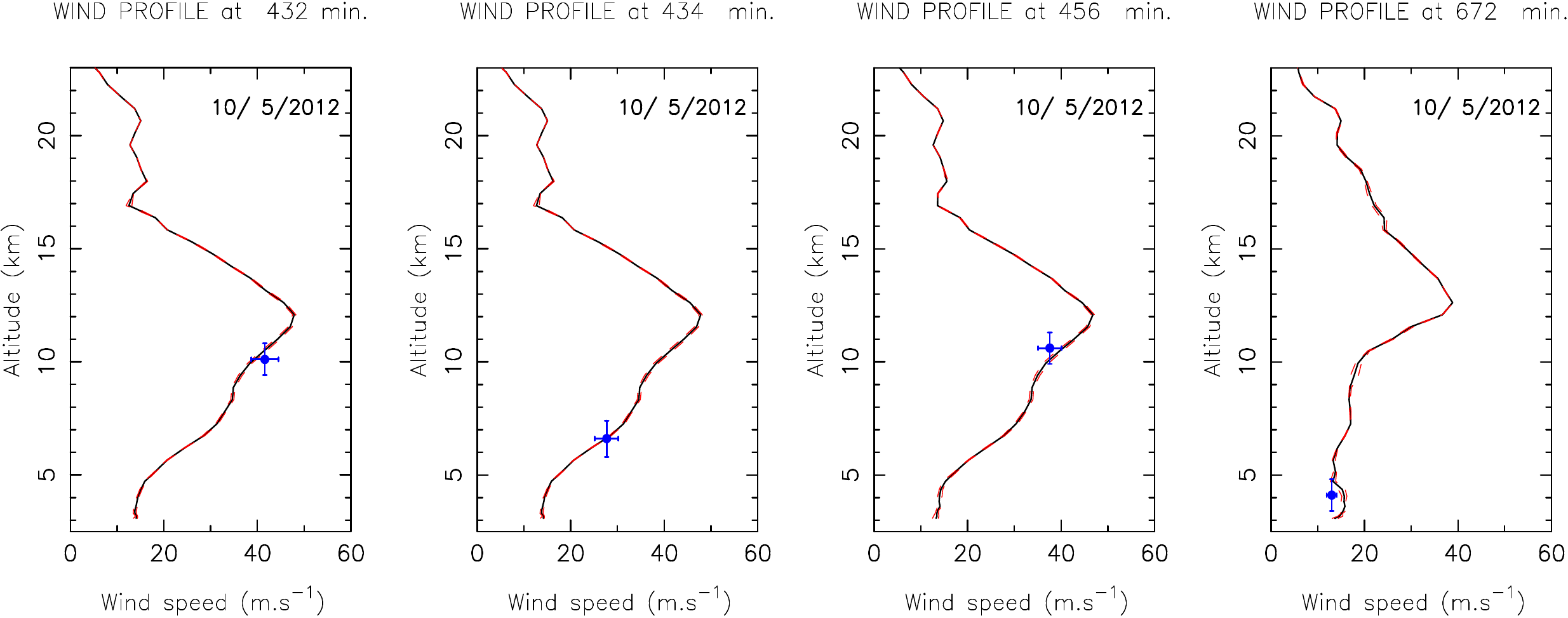}
\end{center}
\caption{\label{fig:10052012} Example of WS as provided by GeMS (blue dots) and by Meso-Nh (vertical profile - black continuum line) during the same night. Red dashed lines are the maximum and minimum variation of the wind speed as simulated by the model in the temporal window $\pm$ 8 min with respect to the closest GeMS measurement. See text for errors bars on x and y-axes. }
\end{figure*}

\begin{figure*} 
\begin{center}
\includegraphics[width=0.9\linewidth]{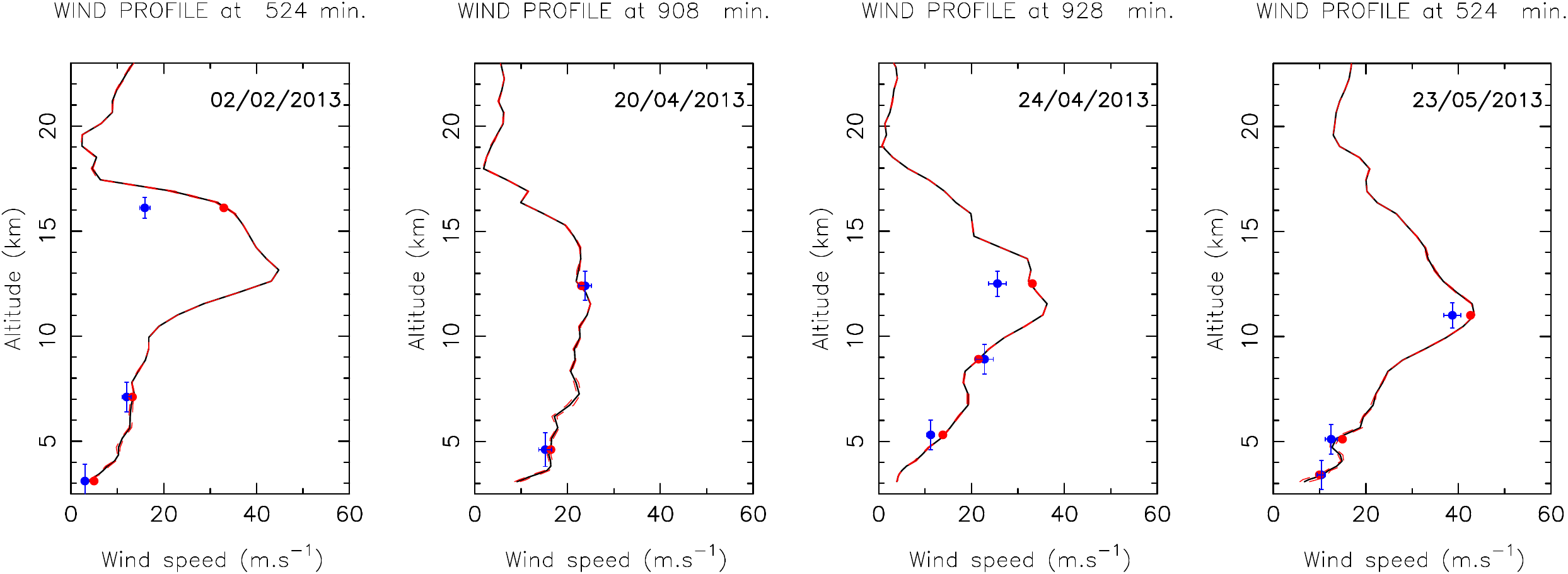}
\end{center}
\caption{\label{fig:varie_wind_speed} 
Examples of WS estimates as provided by GeMS (blue dots) and by Meso-Nh (vertical profile - black continuum line) during a few different nights. See Fig.\ref{fig:10052012} and text of the paper for errors bars on x and y-axes.  Nights in which GeMS detected the largest number of layers are shown. In general, no more than three layers for the WS have been detected by the system. The red dots are displayed on the vertical profile simply to indicate the same heights of the blue dots and facilitate the comparison from a visual point of view.}
\end{figure*}

Fig.\ref{fig:10052012} reports an example related to one night in which is visible a set of estimations of the wind speed as detected by GeMS in different instants of the night (blue dots). In the same figure is reported the vertical WS profile as reconstructed by Meso-Nh in the same instants of GeMS observations. GeMS estimations are clearly in great agreement with estimations obtained with the Meso-Nh model (black continuum line). Concerning to the model estimates, we chose the profiles closest to the time in which observations have been done with respect to the temporal sampling of 2 minutes of Meso-Nh (black continuum line). The maximum and minimum values calculated on a number of profiles selected at $\pm$ 8 minutes with respect to the black line i.e. $\pm$ 4 profiles (dashed red lines) are also displayed. In Fig. \ref{fig:varie_wind_speed} we show the wind speed as reconstructed by the model and as measured by GeMS in a number of nights in which GeMS identified the highest number of turbulent layers where it was possible to perform WS measurements (in the sample of 43 nights it has been detected a maximum number of three layers). In the total sample of 43 nights we can state that the z-axis error bar ($\pm$ error bar) of blue dots (GeMS measurements) is within the $\pm$ [0.5, 0.8] km range and the x-axis error bar is within the $\pm$ [0.5, 3.8] ms$^{-1}$ range. Looking at Fig.\ref{fig:varie_wind_speed} it is possible to note that, in most cases, the WS observed by GeMS and retrieved by Meso-Nh is excellent. Only in one case (2013/02/02), at almost 16 km a.s.l., the WS difference between GeMS and the model is slightly larger. Of course it is not possible to know which of the two (measurement or model estimate) is correct in this specific context. In general the wind shear, particularly that associated to the climatologic features beside the jet-stream at these spatial scales is very well reconstructed by this atmospheric model (see for example \cite{masciadri2013}-Appendix B). Wind shear is associated to dynamic instabilities that are, at the base of the hydrodynamic codes. Moreover, we note that the vertical resolution of the model is almost better than a factor three with respect to the SLODAR at this height. This is more in favour of a correct wind speed from the model. Besides, we observe that, at this height, the wind speed shear is very important. That means that a very small underestimation by GeMS along z should be associated to a much better correlation with the model. In other words, the discrepancy should not be so important. Last but not the least, we can not exclude that, even if less probable, it is the model that slightly overestimates the WS.

\subsection{Wind direction}

\begin{figure*}
\begin{center}
\includegraphics[width = 0.85\linewidth]{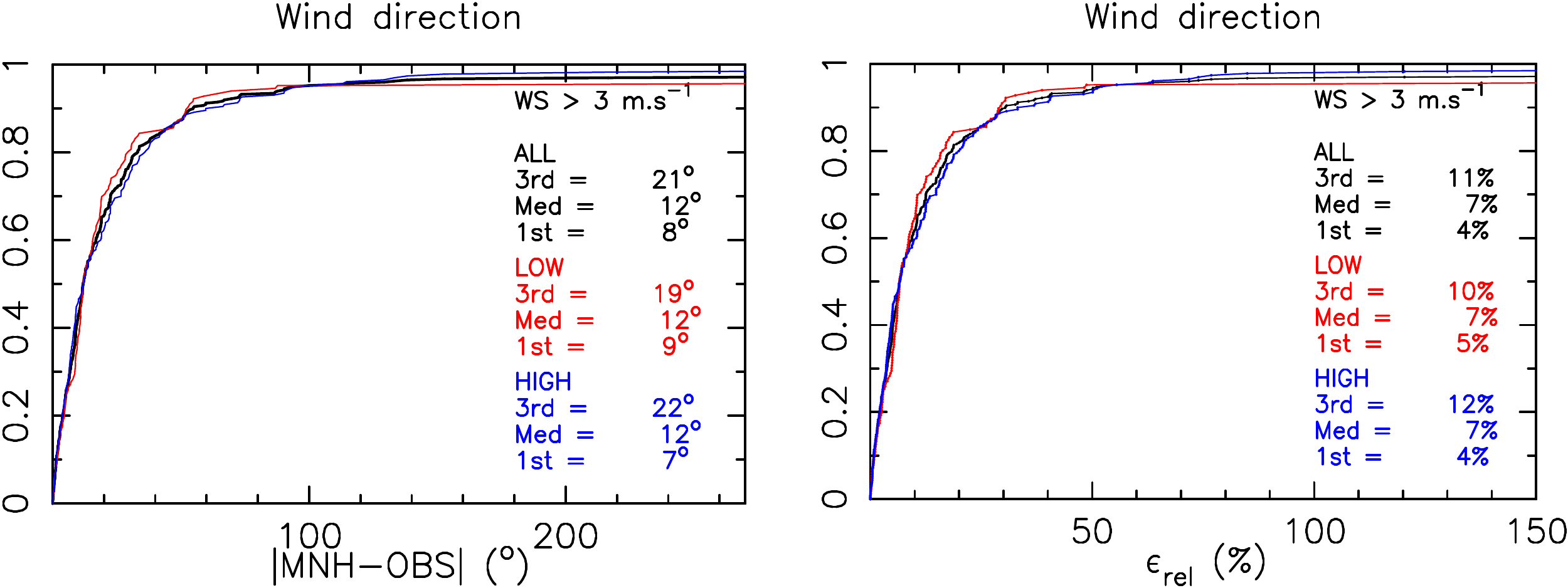}
\end{center}
\caption{\label{fig:cum_dist_dir} Left: Cumulative distribution of the absolute difference of WD estimates retrieved by Meso-Nh and by GeMS (400 points in the sample). Sample contains estimations associated to the [5~km, 18~km] a.s.l. range i.e. the high part of the atmosphere (blue line), the [3~km, 5~km] a.s.l. range i.e. the low part of the atmosphere (red line) and the [3~km, 18~km] a.s.l. range i.e. the total atmosphere (black line). Right: Cumulative distribution of the relative error in the same vertical slabs. There are no GeMS estimates for heights h > 18~km a.s.l.}
\end{figure*}

As we did for the wind speed case, we compared, on the same sample of 43 nights, measured and simulated wind directions on the whole 20 km above the ground. Fig.\ref{fig:cum_dist_dir} reports the cumulative distribution of the absolute difference of the WD as measured by GeMS and estimated by Meso-Nh (as well as the relative error) using the same criteria used for the wind speed. We can observe that the median values for both parameters are excellent with values as small as 12$^{\circ}$ (7\%) in all the [3~km, 5~km] a.s.l., [5~km, 18~km] a.s.l. and [3~km, 18~km] a.s.l. ranges. We divided the atmosphere in this two ranges in order to have two samples that are statistically relevant. For the wind direction not only the median value but also the first and third tertiles remain within excellent values: a maximum value of 22\% for the absolute difference and 12\% for the relative error.

\begin{figure*}
\begin{center}
\includegraphics[width = 0.9\linewidth]{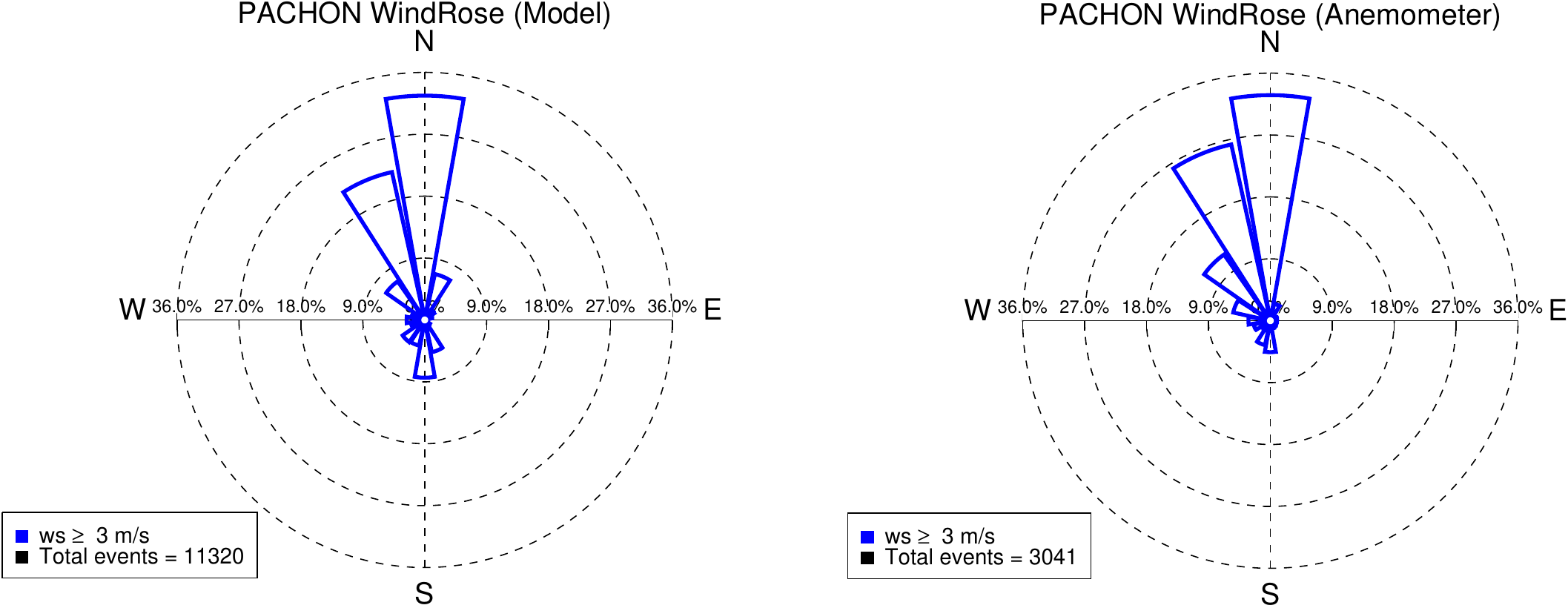}
\end{center}
\caption{\label{fig:winddir_cum} WD histogram retrieved by Meso-Nh (left) and measured by the in situ anemometer (right) on the sample of 43 nights. Observations and simulations having a wind speed weaker than 3 ms$^{-1}$ are filtered out.}
\end{figure*}

Fig.\ref{fig:winddir_cum} shows the wind rose of the wind speed close to the ground for the same sample of 43 nights. The figure reports a histogram of the distribution of the wind direction. Observations are related to the WS and WD measured by the anemometer located in situ at [25-30]~m above the ground at Cerro Pach\'on. Model estimates are related to all the bins included in the same vertical range. In both observations and simulations WD measurements related to the WS weaker than 3 ms$^{-1}$ have been filtered out. Indeed, when the WS is very weak it is meaningless to identify the WD with great accuracy because the uncertainty on the measurement becomes important. As can be seen in Fig.\ref{fig:winddir_cum}, observed and simulated wind direction estimates are in a very good agreement and this is an independent further element that proves the robustness of reference used in this paper i.e. the Meso-Nh model outputs. We remember that results obtained in occasion of the characterization of Cerro Pach\'on atmospheric condition, in perspective of the implementation of the Large Synoptic Survey Telescope (LSST) \citep{els2011}, indicate that the wind blows prevalently from the North-East and North-West in the surface layer above Pach\'on. We conclude that model estimates and measurements from the anemometers, related to our sample of 43 nights, are in excellent agreement with the typical conditions above the Cerro Pach\'on site. All the 400 measurements performed by GeMS are related to heights higher than 45~m a.g.l. and with a lower vertical resolution. It is therefore meaningless to consider the wind rose for the GeMS case. 

\begin{figure*}
\begin{center}
\includegraphics[width = 0.9\linewidth]{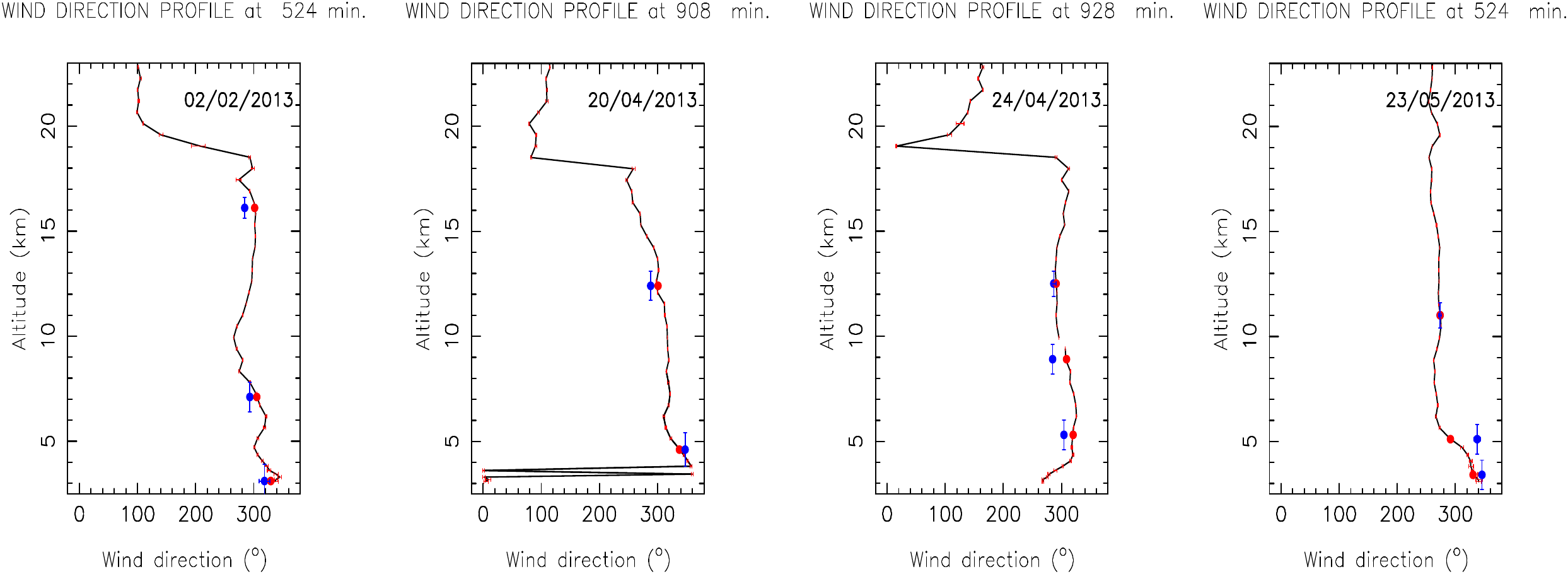}
\end{center}
\caption{\label{fig:windir_indiv} Examples of WD estimates as measured by GeMS (blue dots) and retrieved by Meso-Nh (vertical profile - black continuum line) during different nights. Error bars with respect to the y-axis is the same as that of the wind speed case. Error bars along the abscissa are of the order of a few degrees therefore not visible in the picture. The selected nights are the same as those related to Fig.\ref{fig:varie_wind_speed}. Typically GeMS identifys no more than three layers for the WS. The red dots are displayed on the vertical profile simply to indicate the same heights of the blue dots and facilitate the comparison.}
\end{figure*}

Fig.\ref{fig:windir_indiv} shows the wind direction at different heights as estimated by GeMS and as retrieved by the model for the same set of nights analysed in Fig.\ref{fig:varie_wind_speed}. As already said, these nights correspond to those with the largest number of estimates (typically three) at the same instant. Readers can observe how, in general, the variation of WD increases close to the surface and in the high part of the atmosphere (above the tropopause i.e. above the jet-stream level). This is due to the fact that, in these regions, the wind speed in general decreases and automatically the WD dispersion increases. Looking at Fig.\ref{fig:windir_indiv}, it is possible to observe that, in basically all cases, the agreement model vs. GeMS is extremely good as reflected by the statistical analysis shown in Fig.\ref{fig:cum_dist_dir} giving a medium value of the relative error of 7\%.

\section{Discussion}
\label{discus}

In previous sections GeMS estimates have been compared to independent ones obtained with a non hydrostatic numerical atmospheric model (Meso-NH) that was previously proved to be a robust estimator of WS and WD on the whole column of atmosphere [0-20]~km. For this reason it could be considered as a good reference. The agreement of such a comparison proved, in statistical terms, the validation of the SLODAR technique for wind estimates applied to MCAO systems. This is an important achievement (at our knowledge it the first time such a kind result has been published). However we have to note that the automation of WS and WD measurement based on the SLODAR techniques using a multiple set of sources is not trivial. The principle of detection implies, indeed, the identification of multiple cross-correlation peaks that are sometime not isolated. For a routinely and operational use of a MCAO (or more generally a WFAO) system, the SLODAR technique for WS and WD is, therefore, not very practical. Looking at Fig.\ref{fig:temp_evol1} for example, the reader can retrieve that, during each night, it is possible to isolate only a few number of estimates from GeMS. The temporal coverage is therefore extremely limited. Moreover the automation of such a procedure is not trivial mainly because such a system has been conceived to be an AO system and the measurement of atmospheric parameters has to be seen only as an ancillary output. This does want to be a criticism to the system but just a realistic and pragmatic consideration. It should be much simpler therefore to feed directly the WFAO systems with the Meso-Nh WS and WD estimates. This solution implies the following advantages: (1) to supply information of WS and WD on the 20~km range above the ground, (2) to do that with a temporal sampling of $\sim$ 2 minutes (or even less) and (3) to be available with many hours in advance with respect to the observation. From a practical point of view it should be much simpler to inject in GeMS the predictions of the model with the sequence of couples of (WS,WD) values all along the night with some hours in advance than to recover information on WS and WD from SLODAR technique estimates in real time. The AO system can therefore just read it and use the information as and when it is required. Of course this does not mean that is meaningless to pursue on studies on the measurements done with SLODAR principle of cross-correlation on slopes coming from Shark-Hartmann systems of an MCAO system. However the advantages related to the use of the atmospheric model are a scientific evidence.

Moreover, looking at Fig.\ref{fig:temp_evol1} it is well visible the important spatio-temporal variability of the wind speed on the whole 20~km that might be different in each night. This indicates that, in general, it should be suitable to be able to know the WS and WD with a high frequency in time. The Meso-Nh model can, in this sense, provide a more robust outputs with respect to measurements obtained from GeMS using the SLODAR techniques. The high temporal frequency (2 minutes in this paper) is definitely a very important added value. We note that the frequency of a measurement each two minutes, as provided by the model now, is sufficient to implement the AO control. The fact that it might be, in principle, possible to use a higher frequency (shorter temporal interval) tells us that the predictions provided by Meso-Nh (WS and WD) might be useful in evaluating the validity of the assumption of the 'frozen turbulence' assumption in all those cases dealing with the control techniques as described in the Introduction \citep{gavel2002,poyneer2007,ammons2012,ono2017}. We do not exclude that, under the assumption of a better signal to noise ratio of the MCAO system, measurements and model outputs might be used in a complementary way. That means, for example, that when the measurement done with the MCAO system is missing, the model can replace the information. Anyway, the goal of this paper is not to propose a detailed solution but just to guarantee that an automatic solution can be found because we have reliable model outputs.

GeMS is the first laser assisted MCAO system running on sky and this study proves, for the first time, that the SLODAR technique method for the wind estimate with multiple sources provides reliable results as in agreement with an independent reference previously validated. Besides that, the development of other WFAO systems is in progress for the 8-10 m class telescopes (among these we remind AOF\footnote{AOF: Adaptive Optics Facility} at the Very Large Telescope (\citealt{arsenault2014,madec2016,kuntschner2012}) and LINC-Nirvana\footnote{LINC-NIRVANA: LBT INterferometric Camera - Near InfraRed and Visible Adaptive iNterferometer for Astronomy} at the Large Binocular Telescope \citep{herbst2016}) and others will be implemented on ELT class such as MAORY\footnote{MAORY: Multi-conjugated Adaptive Optics Relay}  \citep{diolaiti2016}, 
the AO system of HARMONI\footnote{HARMONI: High Angular Resolution Monolitic Optical and Near-infrared Integral field spectrograph} (\citealt{neichel2016,thatte2016}) and MOSAIC\footnote{MOSAIC: Multi-Object Spectrograph} (\citealt{hammer2016,morris2016}) at the European-Extremely Large Telescope (E-ELT), NFIRAOS\footnote{NFIRAOS: Narrow Field Infrared Adaptive Optics System} at the Thirty Meter Telescope (TMT) \citep{boyer2014} and GMTAO\footnote{GMTAO: Giant Magellan telescope Adaptive Optics} at the Giant Magellan Telescope (GMT) \citep{bouchez2014}. Besides, a few demonstrators for ELT WFAO systems already seen the first sky (CANARY \citep{gendron2016} at the WIlliam Hershel Telescope and RAVEN \citep{lardiere2014} at the SUBARU Telescope). This study might therefore be taken as a demonstrator for several operational AO and WFAO systems of the forthcoming generation of telescopes.

For completeness we observe that in a couple of papers \citep{neichel2014b,masciadri2016} some preliminary results of these study have been published. In the first one just some preliminary WS comparisons on a very small number of nights has been presented. In the second one we were stuck with serious problems in the comparison of measured and simulated WD estimates. Important discrepancies have been observed and it was not clear the reason of such discrepancy neither if the problem could be referred to GeMS or the model. Recently we could identify the cause of such apparent problem. At that epoch we were unaware of the existence of a flat mirror in the optical path of the GeMS which flips the image plane of the wavefront sensor in the Y direction, making the Y component of the wind vector to appear negative. The problem, that has been now corrected, was, therefore, not in the model neither in GeMS but in the way in which GeMS measurements were read. Moreover, in this paper, new measurements (Gemini anemometer) permitted a precious independent confirmation of results related to WD. We are therefore in the condition not only to validate completely the GeMS instrument, but also to propose a strategy for an automatic WS and WD vertical profiler applied to WFAO systems.

\section{Conclusions}

In this paper we prove, for the first time, the reliability of wind speed and direction measurements obtained with the MCAO system GeMS using a SLODAR technique and multiple laser guide stars sources \citep{guesalaga2014}. The reliability is proved by comparing GeMS estimates of the WS and WD with independent estimates of these parameters obtained with a non-hydrostatic atmospheric mesoscale model (Meso-NH). Such a model has been previously validated by comparing its estimates with 50 different radiosoundings \citep{masciadri2013,masciadri2015}. The level of the agreement between the model and the radio-soundings was so good in statistical terms as well as on each individual profiles that definitely represents an excellent estimation for a cross-comparison with measurements in this study.

By performing the comparison (GeMS vs. model) on a sample of 43 nights (correspondent to around 400 estimates), we find that both, WS and WD estimates provided by GeMS are definitely reliable. The median value of the absolute difference of the GeMS and model WS is equal to 2.5 ms$^{-1}$,  4 ms$^{-1}$ and 3.5 ms$^{-1}$ respectively in the [3~km, 5~km] a.s.l., [5~km, 18~km] a.s.l. and [3~km, 18~km] a.s.l. ranges, that is in the low, high and total atmosphere. Looking at results in terms of relative errors, we find for the three ranges a median value of 26\%, 27\% and 27\% in the three regions. 
Results obtained for the WD are even better: we obtained, for all the three regions, a median value of the absolute difference of the WD from GeMS and from the model equal to 12$^{\circ}$ and a relative error equal to 7\%. The first and third tertiles of WD are within 22\%.  

For an operational application in an operational configuration, we propose, in this paper, to inject the prediction of WS and WD coming from the Meso-Nh model in GeMS. This solution is conceived in alternative to the use of the SLODAR technique. It is, indeed, hard to transform the SLODAR method in a automatic procedure due to its intrinsic limitations (see detailed discussion in Section \ref{discus}). Among these, the main constraint is the difficulty in detecting isolated cross-correlation peaks that are, on the contrary, frequently overlapped. The method we propose presents several advantages (see details in Section \ref{discus}). Besides, the fact that both methods i.e. the SLODAR techniques applied to a MCAO system and the mesoscale model Meso-Nh visibly provide coherent WS and WD estimates, is in favour of a mutual reliability of the two methods. This is certainly useful, particularly for the operational application in which, as we discussed in Session \ref{discus}, the use of the WS and WD provided from the mesoscale model and GeMS can be complementary.

This study can be considered, therefore, a demonstrator for the operational estimation of WS and WD along the 20~km of atmosphere above the ground for whatever WFAO system.

For what concerns the perspectives, it is worth to say that it is known that WFAO systems require not only the knowledge of the WS(h) and WD(h) to optimize their design and to be run operationally but also the knowledge of the optical turbulence stratification (i.e. \CN2(h)). As remembered in Section \ref{sec_obs}, also the optical turbulence stratification can be retrieved using measurements of slopes obtained by a SLODAR technique with some small differences with respect to the calculation of the WS and WD. Starting from the original \citep{cortes2012} study in the last years some diversification and improvements of the original method has been proposed \citep{ono2017,guesalaga2017}. \CN2 vertical profiles can be retrieved also by the Astro-Meso-Nh model \citep{masciadri2017}, but for the optical turbulence it should not be realistic to use the model as a reference. On the contrary \CN2 measurements provided by GeMS might be used, as a reference, to investigate the Astro-Meso-Nh model ability in predicting the optical turbulence above Cerro Pach\'on. This is the next step we envisage to undertake in the next future.

\section*{Acknowledgements}
Based on observations obtained at the Gemini Observatory, which is operated by the Association of Universities for Research in Astronomy, Inc., under a cooperative agreement with the NSF on behalf of the Gemini partnership: the National Science Foundation (United States), the National Research Council (Canada), CONICYT (Chile), Ministerio de Ciencia, Tecnología e Innovación Productiva (Argentina), and Ministério da Ciência, Tecnologia e Inovaçaõ (Brazil). Part of Meso-Nh simulations run on the HPCF cluster of the European Centre for Medium Weather Forecasts (ECMWF) - Project SPITFOT. We acknowledge the Meso-NH Users's support team. AG thanks Conicyt-Chile, grant Fondecyt 1160236. This work has been partially funded by the ANR French program - WASABI.






\newpage





\bsp	
\label{lastpage}
\end{document}